\documentclass{aastex}
\usepackage{emulateapj5}
\usepackage{apjfonts}
\usepackage{float}
\usepackage{graphics}

\usepackage{graphics}

\newcommand{\be}{\begin{equation}}
\newcommand{\ee}{\end{equation}}
\newcommand{\msun}{{M}_{\sun}}

\slugcomment{ApJ Letter, in press}

\begin{document}

%\footnote{Email: qwwu@shao.ac.cn}

\title{Observational Evidence for Young Radio Galaxies are Triggered by Accretion Disk Instability}
\author{Qingwen Wu }
\affil{ Korea Astronomy and Space Science Institute, Daejeon 305348,
Republic of Korea; Email: qwwu@shao.ac.cn}

\begin{abstract}
   The bolometric luminosities and black hole (BH) masses are
   estimated by various methods for a sample of young radio galaxies with
   known ages. We find that the ages are positively
   correlated with the bolometric luminosities in these young radio
   galaxies, which is consistent with theoretical prediction
   based on radiation pressure instability of accretion disk in Czerny et al.
    The ages of young radio galaxies are also found to be
   consistent with the theoretical durations of outbursts in
   BH mass and accretion rate (in Eddington unit) plane, where the outbursts are assumed
   to be triggered by the radiation pressure instabilities. Our results provide
   the observational evidence for the radiation pressure instability, which
    causes limit-cycle behavior, as a physical mechanism that may
    be responsible for these short-lived young radio galaxies.

\end{abstract}


\section{Introduction}
Gigahertz-peaked spectrum (GPS: projected linear size $D\lesssim1$
kpc) and compact steep-spectrum (CSS: projected linear size
$D\lesssim20$ kpc) radio sources constitute a large fraction
($\sim40\%$) of the powerful radio source population. Their radio
spectra are simple and convex with peaks close to 1 GHz and 100 MHz
for GPS and CSS sources respectively (see O'Dea 1998 for a review).
Two basic models have been proposed to explain the CSS/GPS
phenomena: (1) \emph{youth scenario}: the radio sources are compact
because they are young, and the jets will expand into large-scale
`classical' lobes during the course of their life (e.g., Fanti et
al. 1995); (2) \emph{frustration scenario}: the radio emitting
plasma is permanently confined to the region within the host galaxy
by a cocoon of dense gas and dust for their entire life (e.g., van
Breugel 1984). Both kinematic ages estimated from proper motions of
hot spots (e.g., Giroletti \& Polatidis 2009 and references therein)
and radiative ages deduced by synchrotron theory based on the radio
spectra of lobes (e.g., Murgia 1999 and reference therein) provide
values from $\sim10^2$ to $\sim10^5$ years, which strongly support
the \emph{youth scenario}.

 Duty cycle is one of key problems in understanding the evolution of
active galactic nuclei (AGN, e.g., Wang et al. 2009, and references
therein). The extended radio emission in galaxies and quasars
provides us with an opportunity to probe their history through the
ages derived from the structural and spectral information of their
radio lobes. Both the young compact symmetric object (CSO) with a
much older extended relic emission (e.g., Owsianik et al. 1998) and
the `double-double' radio source which consists of a new lobe-pair
near the core and a more distant faded lobe-pair (e.g., Lara et al.
1999; Schoenmakers et al. 2000) can be explained in terms of
recurrent activity in their nuclei. These possibly re-born radio
sources may be intermittent on timescales of $\sim10^{4}-10^{5}$
years (e.g., Reynolds \& Begelman 1997).

A standard, geometrically thin accretion disk (SSD, Shakura \&
Sunyaev 1973) is known to be unstable in the innermost, radiation
pressure dominated regions when accretion rate is high (e.g.,
Lightman \& Eardley 1974). The thermally unstable inner region will
possibly lead to the so-called limit-cycle behavior, which has been
confirmed by numerical simulations of the time-dependent disk (e.g.,
Honma et al. 1991; Li et al. 2007).  On the observational side, a
perfect candidate of black hole (BH) X-ray binary GRS 1915+105 has
been known to show the theoretically predicted limit-cyclic
luminosity variations (e.g., Belloni et al. 1997; Nayakshin et al.
2000; Janiuk et al. 2002; Merloni \& Nayakshin 2006). Czerny et al.
(2009) found that the outburst durations are generally from
$\sim10^{2}$ to $\sim10^{5}$ years for BH mass $M_{\rm
BH}=10^{7}-10^{10}\msun$ and viscosity parameter $\alpha=0.02-0.2$,
which are roughly consistent with the observed ages of young radio
galaxies, where the viscosity law with a geometrical mean between
the gas and the total pressure for scaling of the stress, i.e.,
$T_{r\phi}\propto(P_{\rm gas}P_{\rm tot})^{1/2}$, is adopted when
modeling the time evolution of SSD under the radiation pressure
instability. An important empirical correlation between the outburst
duration, $T_{\rm burst}$, bolometric luminosity, $L_{\rm bol}$, and
viscosity parameter, $\alpha$, is given by Czerny et al. (2009), \be
\log\left(\frac{T_{\rm burst}}{\rm yr}\right)\simeq1.25\log
\left(\frac{L_{\rm bol}}{\rm
erg/s}\right)+0.38\log\left(\frac{\alpha}{0.02}\right)-53.6.\ee

Wu (2009) found that most of young radio galaxies have relatively
high accretion rates with average Eddington ratio $<\log L_{\rm
bol}/L_{\rm Edd}>$= -0.56 for a sample of 51 sources (32 CSS + 19
GPS), where $L_{\rm Edd}=1.38\times10^{38}M_{\rm BH}/\msun$ is
Eddington luminosity. The accretion disk in these high Eddington
ratio CSS/GPS sources may have suffered the radiation pressure
instability, which provide a possibility to explain their
short-lived activities. In this \emph{Letter}, we estimate the
bolometric luminosities and BH masses for a sample of young radio
galaxies with known ages. We then test whether the ages of young
radio galaxies are consistent with the durations of outbursts or
not, where the outbursts are assumed to be triggered by the
radiation pressure instabilities of accretion disk (e.g., Czerny et
al. 2009). Throughout this paper, we assume the following cosmology:
$H_{0}=70\ \rm km\ s^{-1} Mpc^{-1}$, $\Omega_{0}=0.3$ and
$\Omega_{\Lambda}=0.7$.

\section{The sample}
  We compile a sample of 37 young radio sources from the data available
  in the literature. The kinematic ages are mainly selected
  from Giroletti \& Polatidis (2009, and references therein), while
  the ages estimated from synchrotron cooling time are mainly taken from
  Murgia et al. (1999). We give the lower limits of kinematic age for four
  sources (0134+329, 0538+498, 1458+718,
  2249+185) assuming the upper limit for the
  separation of hot spots to be the light speed $c$ in this work, since their radiative ages may be much shorter
   than the actual source ages in these jets or hot spots dominated sources (see Murgia et al.
  1999 for more details). The kinematic age of CSS source 1328+307 (3C 286)
  is estimated in the present work from its projected liner size and jet propagation speed
  $\simeq0.1c$ (An et al. 2004). Several other individual sources
  with known ages are also selected from the literature. In total, our sample consists of
 14 GPS and 23 CSS sources (see Table 1 for more details).

\section{Bolometric luminosities and BH masses}

   The bolometric luminosity of AGN is sometimes approximated from the
  optical luminosity, since the integration of spectral energy
  distribution (SED) is usually hampered by the lack of
  multi-wavelength observational data that spans many decades in
  wavelength. It is found that the bolometric luminosity is roughly 9 times
  the optical luminosity, $L_{\rm bol}=9 L_{5100}$, where $L_{5100}=\lambda L_{\lambda}(5100\rm \AA)$ (Kaspi et al.
  2000). It should be noted that the monochromatic optical luminosity cannot be
  used to derive the bolometric luminosity if the optical luminosity is contaminated by the jet emission
 in radio loud AGN (e.g., BL Lacs) or obscured by the putative
 torus. The broad line emission is believed to be illuminated by the ionizing luminosity from the accretion disk,
  and, therefore, it can be used as a good indicator of thermal
  optical emission. We estimate the bolometric luminosities for 9 sources
  from their broad line luminosities (e.g., $\rm H\beta$, C IV)
  using the empirical correlations between the luminosities of different emission
  lines and bolometric luminosities derived from a sample of radio-quiet AGN in Wu
  (2009). The bolometric luminosities of 2 sources fitted from
  their SEDs are selected directly (2342+821, 1458+718, Woo \& Urry 2002). The bolometric luminosity of 1 source
   have been corrected for beaming effect is also included (1413+135, Xie et al. 2006) .
   The bolometric luminosities of 4 optically Type 1 CSS sources are estimated
   from their optical luminosities with $L_{\rm bol}=9 L_{5100}$,
   since that, normally, the steep-spectrum sources do not suffer
   strong contamination from the Doppler-boosted emission of the jet
    due to the possible large inclination angles.

  In the standard model, narrow emission lines (e.g., [O III]) are also believed to be excited through
 photoionization by the UV continuum produced by the accretion disk (e.g., in radio-quiet
 AGN). However, it is more complicated in young radio galaxies,
 since shocks caused by the interaction between the jet and interstellar
 medium (ISM) may also be important for the morphology, kinematics, and the excitation of
narrow emission lines (e.g., Labiano et al. 2005; Holt et al. 2006).
Moy \& Rocca-Volmerange (2002) proposed that the shock ionization
may be only important for sources with intermediate radio sizes (2
kpc$<$$\it D$$<$150 kpc, e.g., CSS sources), while AGN
photoionization is dominant in the most compact sources ($D$$<$2
kpc, e.g., GPS sources) and the extended sources with $D$$>$150 kpc.
Moreover, the jet sizes in all GPS sources of our sample are less
than 0.2 kpc (see Table 1), which are much less than the kpc-scale
of narrow line regions. This also support that the shock
contribution to narrow lines from the jet-cloud interaction may be
not important and, therefore, narrow line emission may still be a
good indicator of central thermal emission in these GPS sources. To
derive the bolometric luminosities for more sources, we investigate
the relation between [O III] luminosities and bolometric
luminosities for a sample of radio-quiet Type 1 AGN, which include
20 Seyferts and 23 low red-shift QSOs. The Seyferts are selected
from Kaspi et al. (2000), where the [O III] luminosities are chosen
from Kraemer et al. (2004), and the QSOs are selected from
Alonso-Herrero et al. (1997). To be consistent with other methods,
we calculate the bolometric luminosities from $L_{\rm bol}=9
L_{5100}$ for these radio-quiet AGN. The relation between the
bolometric luminosities and [O III] luminosities is shown in Figure
1, and its linear regression is, \be \log L_{\rm
bol}=0.95\pm0.08\log L_{\rm [O\ III]}+5.39\pm3.17. \ee We find that
bolometric correction factors, $L_{\rm bol}/L_{\rm [O\ III]}$, are
about 2000 and 3400 for 20 Seyferts and 23 QSOs respectively. The
bolometric luminosities for 12 GPS and 9 CSS sources were calculated
with Eq. (2) from their [O III] luminosities. One should keep in
mind that the bolometric luminosities of 9 CSS sources may be
over-estimated due to the possible shock contribution to [O III]
luminosities (see discussions in Sect. 5).

\figurenum{1}
\centerline{\includegraphics[angle=0,width=9.0cm]{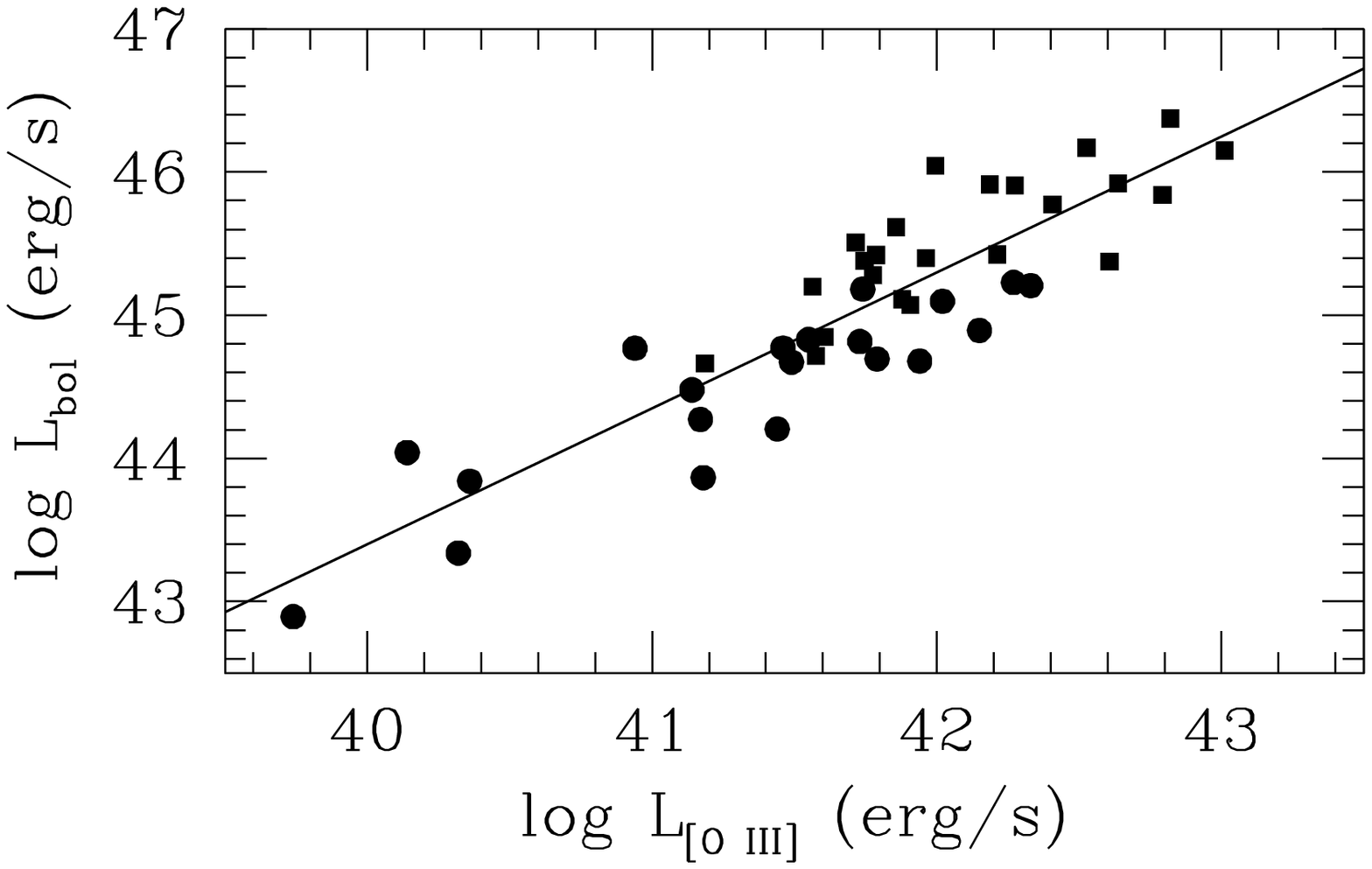}}
\figcaption{\footnotesize The correlation between bolometric
luminosities and [O III] luminosities for a sample of radio-quiet
Type 1 AGN. The circles and squares are correspond to 20 Seyferts
and 23 QSOs respectively. The solid line is the best fitting.
\label{fig1}} \centerline{}

  Wu (2009) have estimated BH masses for a sample of 65 young radio
galaxies from different methods. We estimate BH masses of other 14
young radio galaxies, which are not included in Wu (2009), from the
relation between host galaxy absolute magnitude $M_{R}$ in $R$-band
and BH mass proposed by McLure et al. (2004), \be \log_{10}(M_{\rm
BH}/\msun)=-0.5M_{R}-2.74.\ee

\section{Results}

      We estimate the bolometric luminosities for 37 young radio sources by various methods.
    The relation between ages and bolometric luminosities for young
     radio galaxies is presented in Figure 2, where the bolometric
   luminosities of CSS sources calculated from [O III] luminosities are marked with extra open
   squares. We find that the ages are positively
   correlated with the bolometric luminosities in these young radio galaxies, which is consistent with
   the prediction of radiation pressure instability in accretion disk (Czerny et al.
   2009). We plot the dashed lines corresponding to Eq. (1) with viscosity
   parameters $\alpha=$$10^{-4}$ and 1 in Figure 2. We
   find that most sources are locate to the left of the line with
    $\alpha=1$ ($L_{\rm bol}$-$T_{\rm burst}$), where the $\alpha=1$ line gives
  an upper limit of the theoretical outburst duration.  The relation between
   bolometric luminosity and half of the outburst duration
   with $\alpha=0.1$ is also presented in Figure 2 ($L_{\rm bol}$-$T_{\rm burst}/2$, solid line), which is
    converted from the Eq. (1). We find that the real ages of
    sources in our sample are roughly consistent with the theoretical
    prediction of $T_{\rm burst}/2$ with $\alpha\simeq0.1$, since that
    sources should have the real ages between 0 and $T_{\rm burst}$ (e.g., 0$<$age$<$$T_{\rm
    burst}$), and, on the average, the real ages should
    correspond to $T_{\rm burst}/2$ with some dispersion around this value.

\figurenum{2}
\centerline{\includegraphics[angle=0,width=9.0cm]{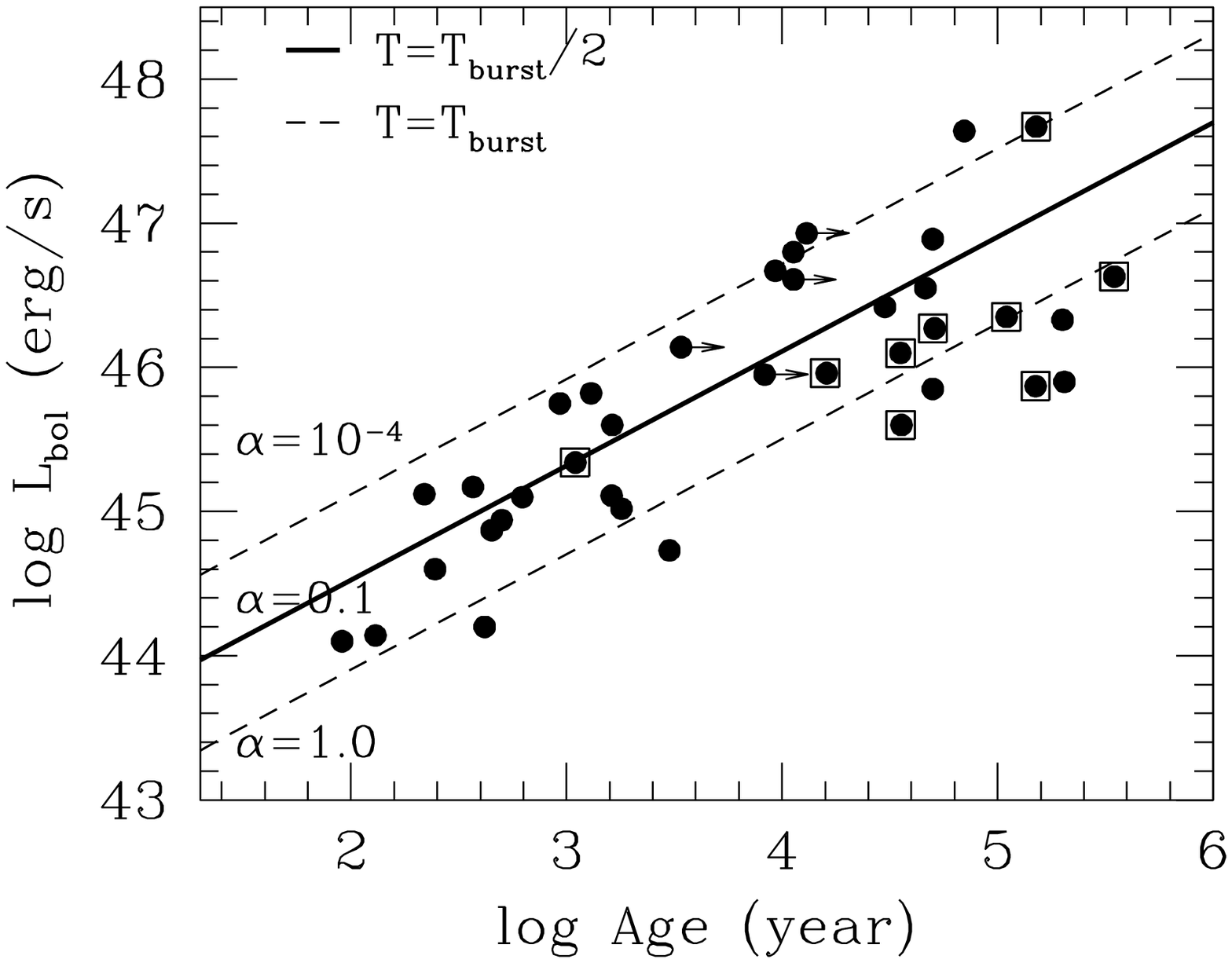}}
\figcaption{\footnotesize The correlation between ages and
bolometric luminosities for young radio galaxies. The dashed lines
are theoretical predictions from the disk instability model for
$L_{\rm bol}-T_{\rm burst}$ with $\alpha=$$10^{-4}$ (upper) and 1
(lower) respectively, while the solid line is for $L_{\rm
bol}-T_{\rm burst}/2$ with $\alpha=$0.1. The bolometric luminosities
of CSS sources calculated from [O III] luminosities are marked with
extra open squares. \label{fig2}} \centerline{}

\figurenum{3}
\centerline{\includegraphics[angle=0,width=9.0cm]{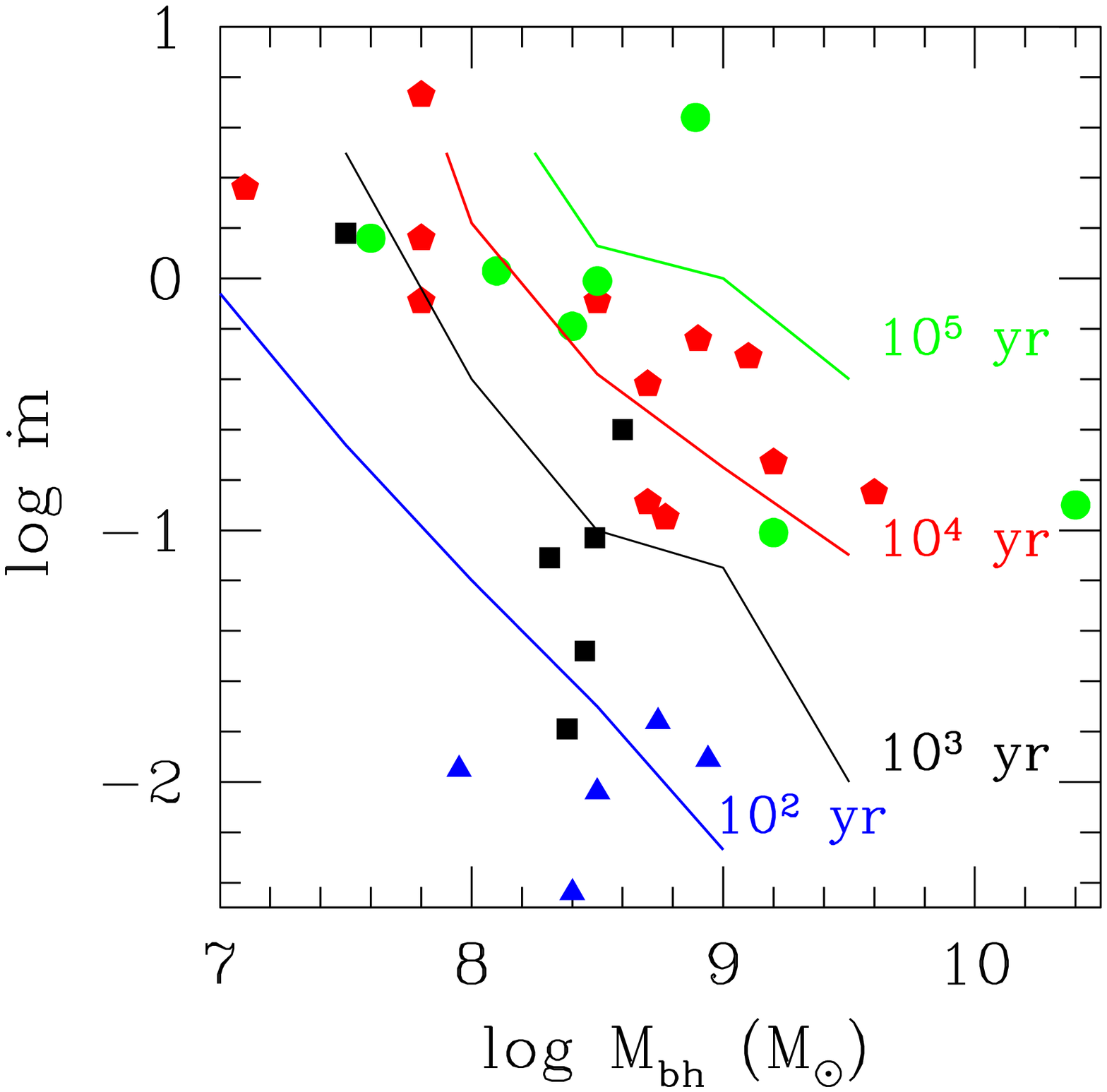}}
\figcaption{\footnotesize Comparison of the ages for young radio
galaxies with the theoretical durations of outbursts triggered by
disk instability in $M_{\rm BH}-\dot{m}$ plane. The triangles,
squares, pentagons, and circles correspond to the age in bins of
[0-$5\times10^{2}$], [$5\times10^{2}$-$5\times10^{3}$],
[$5\times10^{3}$-$5\times10^{4}$], and
[$5\times10^{4}$-$5\times10^{5}$]yrs respectively. The solid lines
are the half of theoretical outburst durations predicted by the disk
instability model with $\alpha=0.1$, which are converted from the
case of $\alpha=0.2$ in Czerny et al. (2009, Fig. 5) using the
scaling relation of Eq. (1).  \label{fig3}} \centerline{}

The BH masses are calculated from the luminosities of host galaxy,
or collected from the literature. The Eddington ratios,
$\dot{m}=L_{\rm bol}/L_{\rm Edd}$, were computed from the estimated
bolometric luminosities and BH masses. We then compare the ages of
young radio sources with the half of outburst durations of in BH
mass-accretion rate (in Eddington unit, $M_{\rm BH}-\dot{m}$) plane,
where the outbursts are assumed to be triggered by the radiation
pressure instabilities of accretion disk. To do so, we convert the
contour map of $\alpha=0.2$ in Czerny et al. (2009, Fig. 5) to the
case of $T_{\rm burst}/2$ with $\alpha=0.1$ using the scaling
relation of Eq. (1), since we find that most of the young radio
sources can, on the average, be described by $\alpha\simeq0.1$ in
the $L_{\rm bol}$-age relation. The ages of young radio sources are
divided into four bins as [0-$5\times10^{2}$],
[$5\times10^{2}$-$5\times10^{3}$],
[$5\times10^{3}$-$5\times10^{4}$], and
[$5\times10^{4}$-$5\times10^{5}$]yrs according to the theoretical
lines of $10^{2}$, $10^{3}$, $10^{4}$ and $10^{5}$yrs of Czerny et
al. (2009) respectively. The results are shown in Figure 3. We find
that, on the average, the ages of young radio sources are also
roughly consistent with the half of outburst durations with
$\alpha\simeq0.1$ in $M_{\rm BH}-\dot{m}$ plane, even with
substantial scatter. In particular, there are 12 sources in the bin
of [$5\times10^{3}$-$5\times10^{4}$]yrs (red-pentagons), which
coincide with the theoretical prediction of $10^{4}$yrs (red-solid
line) very well.

\section{Summary and Discussion}

   We estimate bolometric luminosities and BH masses for a
   sample of young radio galaxies with known ages. We find a
   positive correlation between the ages and bolometric luminosities
   in these young radio sources (see Fig. 2). This positive correlation
   is roughly consistent with the theoretical prediction based on
   the radiation pressure instability of accretion disk in Czerny et al.
   (2009). We find that the real ages of most sources are shorter than
   the upper limit of outburst duration with $\alpha=1$, and, on the average,
   correspond to the half of outburst durations with $\alpha\simeq0.1$ (see Fig.
   2). We further compare the ages with the half of outburst durations
   in $M_{\rm BH}-\dot{m}$ plane, and find
   that the observational data is also roughly consistent with the theoretical
   prediction with $\alpha\simeq0.1$ (see Fig. 3). Therefore, our results provide
   observational evidence for the radiation pressure instability as a physical mechanism that may
   also be responsible for these short-lived young radio galaxies,
   beside the perfect application to the brightest Galactic X-ray binary GRS 1915+105.

   The bolometric luminosities for young radio galaxies
   are estimated from their SED fittings, optical luminosities, broad
   line luminosities, or narrow line luminosities. In these
   methods, the most controversial one is that derived from the [O III] luminosity,
   due to the possible shock contribution caused by the jet-ISM interaction.
    However, the shock contribution may be not important in the GPS
    sources as suggested by the ionization diagnostics of different
    narrow emission lines (e.g., Moy \& Rocca-Volmerange 2002). Wu (2009) also found that
    accretion activities may still play an important role in shaping the kinematics
    of [O III] narrow line in young radio galaxies. The
    bolometric luminosity calculated from [O III] luminosity is
    $\log L_{\rm bol}(\rm [O\ III])=45.12$ for the GPS source 1404+286,
    which is consistent with that calculated from the broad
    line luminosity $\log L_{\rm bol}(\rm H\beta)=45.19$ very well,
    where the [O III] luminosity is selected from Marziani et al. (2003).
    This also support that the shock contribution to [O III]
    luminosity may not be important, and the [O III]
    luminosity can still be regarded as a good indicator of central thermal
    emission in these small radio-sized GPS sources. Labiano et al.
    (2005) found that the contribution to narrow line
    luminosity from the shocked gas is between 30 and 70\% for CSS
    source 3C 303.1, and this ratio is even lower in
    other two CSS sources (3C 67 and 3C 277.1), based on the ionization
    diagnostics of different narrow emission lines. Therefore, the bolometric luminosities of 9 CSS
    sources calculated from the [O III] luminosities may not be
    over-estimated too much, which will not change our main
    conclusion.

  The limit-cycle behavior triggered by the radiation pressure
  instability of accretion disk provide a good
  explanation to the luminosity variations of X-ray binary GRS 1915+105 (e.g., Nayakshin et al.
  2000; Janiuk et al. 2002; Merloni \& Nayakshin 2006). Czerny et al.
  (2009) proposed that the young radio galaxies may have the similar
  origin, and the unstable disk caused by radiation pressure may be the physical
  reason for their short-lived radio activities.  We note that the unique
  limit-cycle variability only appears in the brightest X-ray binary GRS
  1915+105 when it radiates at high luminosity (e.g., Belloni et al.
  1997). Wu (2009) found that most of the young radio galaxies
  have relatively high Eddington ratios, and the accretion disk may
  also have suffered the radiation pressure instabilities in these
  young radio sources. We collect a sample of 37 young radio
  galaxies, and explore whether these young radio sources are consistent with
  the predictions of disk instability model in Czerny et al. (2009) or
  not.  We test this in both the $L_{\rm bol}$-age relation and the
  ages in $M_{\rm BH}-\dot{m}$ plane.
  We find that the real ages of most sources are, indeed, less than the upper limit
  of the theoretical outburst durations with $\alpha=1$ (see Fig. 2).
  The viscosity parameter, $\alpha$, should be larger than
  $10^{-4}$, since that the real ages of most sources are longer than
  its predicted outburst durations (see Fig. 2).
  We find that the real ages of these young radio sources are
   roughly consistent with the model predictions
  with $\alpha\simeq0.1$, if we assume the real ages, on the average, should
  correspond to the half of outburst durations (see Fig. 2).
  The real ages of the young radio sources are also
  found to be consistent with the half of outburst durations with $\alpha\simeq0.1$ in $M_{\rm
  BH}-\dot{m}$ plane (see Fig. 3). It should be noted that
  there are still uncertainties in estimating BH masses, bolometric
  luminosities and ages in these young radio galaxies. Moreover,
  the simplified disk model was used to derive the Eq. (1), where corona
  and jet are neglected (see Czerny et al. 2009 for more details).
  With all these uncertainties in mind, it seems that the
  observational data of young radio galaxies are in good agreement
  with the theoretical predictions of disk instability model.  We
  note that the disk instability model also predicts a complex behavior of the
  jet activities depending on the accretion rate and BH mass: (1) continuous
  jet at low accretion rate where the accretion disk is stable
  (e.g., $\dot{m}\lesssim10^{-2}$, low luminosity AGN); (2) intermittent compact jet
  with short outburst duration ($\lesssim 10^{4}$yr) at high
  accretion rate, where the disk is unstable and the short-lived outburst of central activity
  cannot escape from host galaxy (e.g., $\dot{m}\gtrsim10^{-2}$, GPS and CSO sources);
  (3) intermittent jet with long outburst duration ($\gtrsim 10^{4}$yr)
    at high accretion rate (e.g., $\dot{m}\gtrsim10^{-2}$, CSS sources),
   where the disk is unstable and the jet is able to
   escape from the host galaxy and possibly grow to large-scale
   radio source (e.g., FR I/FR II, see Czerny et al. 2009 for more details).

\acknowledgments
   We thank the anonymous referee for constructive comments that helped to improve the paper.
   We also thank Czerny, B., Cao, X. W., Zheng, S.-M., Lu, J.-F.,
   Pariev, V. and Das, S. for their careful reading of the manuscript, valuable
   comments and suggestions. This work is partly supported by the NSFC (grants number 10703009).

\clearpage
\begin{table*}[t]
\footnotesize %\scriptsize
 \begin{flushleft}
  \centerline{\bf Table 1. The data of young radio sources.}
  \begin{tabular}{lcccccccccccc}\hline
Name  & Others & $z$ & Size &  Age & Method & Refs.$^{a}$ & $L_{\rm
bol}$ & Method & Refs.$^{b}$ &  $M_{\rm bh}$ & Refs.$^{c}$ & $L_{\rm
bol}/L_{\rm Edd}$  \\
             &           &         & kpc  & [$10^{3}$yrs]&  &   & $\log$ (ergs/s) & &    &$\log (\msun$) & &$\log$  \\
 \hline
             &           &         &       &         &     &  \emph{\textbf{GPS}}      &        &          &       &      &       &   \\
   0108+388  &  ...      &  0.669  & 0.023 &   0.40  & kin &  G09   & 44.20  & L(O III)  & L96  & 7.9  & 21.6$^{1}$  &   -1.82 \\
   0710+439  &  ...      &  0.518  & 0.088 &   0.93  & kin &  P03   & 45.75  & L(O III)  & L96  & 8.4  & 20.0$^{1}$  &   -0.75 \\
   1031+567  &  ...      &  0.450  & 0.109 &   1.84  & kin &  T00   & 45.02  & L(O III)  & L96  & 8.1  & 20.2$^{1}$  &   -1.23 \\
   1358+624  & 4C 62.22  &  0.431  & 0.160 &   2.4   & kin &  V06   & 45.10  & L(O III)  & L96  & 8.2  & 19.8$^{1}$  &   -1.27 \\
   1404+286  & OQ 208    &  0.077  & 0.007 &   0.22  & kin &  L07   & 45.19  &L(H$\beta$)& M03  & 8.7  & 14.6$^{1}$  &   -1.69 \\
   1934-638  &  ...      &  0.183  & 0.085 &   1.60  & kin &  G09   & 45.60  & L(O III)  & X99  & 8.5  & 17.1$^{1}$  &   -1.03 \\
   2021+614  &  ...      &  0.227  & 0.016 &   0.37  & kin &  T00   & 45.17  & L(O III)  & L96  & 8.9  & 16.8$^{1}$  &   -1.91 \\
   2352+495  &  ...      &  0.238  & 0.117 &   3.0   & kin &  P03   & 44.73  & L(O III)  & L96  & 8.4  & 18.0$^{1}$  &   -1.79 \\
   J1111+1955&  ...      &  0.299  & 0.070 &   1.62  & kin &  G05   & 45.11  & L(O III)  & P00  & 8.5  & 18.4$^{1}$  &   -1.48 \\
   0116+31   & 4C 31.04  &  0.060  & 0.071 &   0.50  & kin &  G09   & 44.94  & L(O III)  & X99  &  ... & ...         &     ... \\
   1718-649  & NGC 6328  &  0.014  & 0.002 &   0.09  & kin &  G09   & 44.10  & L(O III)  & X99  & 8.4  & 11.4$^{2}$  &   -2.44 \\
   0316+413  & 3C 84     &  0.018  & 0.015 &   0.24  & kin &  N09   & 44.60  & L(O III)  & H01  & 8.5  & W02         &   -2.04 \\
   1413+135  &  ...      &  0.247  & 0.031 &   0.13  & kin &  G05   & 44.14  & Fitting   & X06  & 8.0  & 18.9$^{1}$  &   -1.95 \\
   0035+227  &  ...      &  0.096  & 0.022 &   0.45  & kin &  G09   & 44.87  & L(O III)  & G04  &  ... & ...         &     ... \\
             &           &         &       &         &     &  \emph{\textbf{CSS}}    &    &          &       &      &       &   \\
   0221+276  & 3C 67     &  0.310  & 6.8   &   51    & syn &  M99   & 46.27  & L(O III)  & H03  & 8.1  & W09         &    0.03 \\
   0404+769  & 4C 76.03  &  0.599  & 0.53  &   1.1   & syn &  M99   & 45.34  & L(O III)  & L96  & 8.3  & 20.3$^{1}$  &   -1.11 \\
   0740+380  & 3C 186    &  1.067  & 8.2   &   113   & syn &  M99   & 46.80  &L(H$\beta$)& W09  & 8.9  & W09         &   -0.30 \\
   1019+222  & 3C 241    &  1.617  & 2.8   &   9.3   & syn &  M99   & 46.67  &L(H$\beta$)& W09  & 7.8  & W09         &    0.56 \\
   1203+645  & 3C 268.3  &  0.371  & 3.9   &   35.4  & syn &  M99   & 46.10  & L(O III)  & H03  & 7.8  & W09         &    0.16 \\
   1250+568  & 3C 277.1  &  0.320  & 4.4   &   204   & syn &  M99   & 45.90  &L(H$\beta$)& W09  & 7.6  & W09         &   -0.14 \\
   1416+067  & 3C 298    &  1.437  & 9.1   &   70    & syn &  F02   & 47.64  & L5100     & W09  & 10.4 & W09         &   -0.83 \\
   1443+77   & 3C 303.1  &  0.267  & 5.0   &   110   & syn &  M99   & 46.35  & L(O III)  & H03  &  8.4 & W09         &   -0.19 \\
   1447+77   & 3C 305.1  &  1.132  & 9.0   &   151   & syn &  M99   & 47.67  & L(O III)  & A00  &  8.9 & 21.0$^{2}$  &    0.64 \\
   2252+12   & 3C 455    &  0.543  & 13.7  &   348   & syn &  M99   & 46.63  & L(O III)  & X99  &  8.5 & W09         &   -0.01 \\
   2342+821  &  ...      &  0.735  & 0.66  &   1.3   & syn &  M99   & 45.82  & Fitting   & W02  &  7.5 & W09         &    0.18 \\
   0127+233  & 3C 43     &  1.459  & 9.4   &   200   & syn &  F02   & 46.33  & L5100     & W09  &  9.2 & W09         &   -1.01 \\
   0134+329  & 3C 48     &  0.367  & 2.1   & $>$10   & kin &  A1    & 46.61  &L(H$\beta$)& W09  &  9.2 & W09         &   -0.30 \\
   0518+165  & 3C 138    &  0.759  & 2.9   &   50    & kin &  A93   & 46.42  &L(H$\beta$)& W09  &  8.7 & W09         &   -0.42 \\
   0538+498  & 3C147     &  0.545  & 2.4   & $>$3.9  & kin &  A1    & 45.95  &L(H$\beta$)& W09  &  8.7 & W09         &   -0.90 \\
   0758+143  & 3C 190    &  1.195  & 14.1  &   50    & syn &  K97   & 45.85  &L(H$\beta$)& W09  &  7.8 & W09         &   -0.13 \\
   1328+307  & 3C 286    &  0.849  & 14.2  &   462   & kin &  A2    & 46.55  & L5100     & W09  &  8.5 & W09         &   -0.09 \\
   1328+254  & 3C 287    &  1.055  & 0.4   &$55\pm45$& syn &  T01   & 46.89  & L5100     & W09  &  9.6 & W09         &   -0.85 \\
   1458+718  & 3C 309.1  &  0.905  & 7.8   & $>$13   & kin &  A1    & 46.93  & Fitting   & W02  &  9.1 & W09         &   -0.31 \\
   2249+185  & 3C 454    &  1.757  & 2.1   & $>$3.4  & kin &  A1    & 46.14  & L (CIV)   & W09  &  8.6 & W09         &   -0.58 \\
   0345+337  & 3C 93.1   &  0.243  & 1.2   &   35.8  & syn &  M99   & 45.60  & L(O III)  & X99  &  7.1 & W09         &    0.36 \\
   1517+204  & 3C 318    &  1.574  & 7.8   &   150   & syn &  T92   & 45.87  & L(O III)  & H03  &  ... & ...         &    ...  \\
   0809+404  & 4C 40.2   &  0.551  & 4.8   &   100   & syn &  K06   & 45.96  & L(O III)  & Z03  &  8.8 & 19.4$^{2}$  &   -0.95 \\
\hline
\end{tabular}
 \tablenotetext{a}{References for the radio sizes, the ages of these young radio sources, and the methods for deriving the ages.}
\tablenotetext{b}{References for the data used to derive the
bolometric luminosities.} \tablenotetext{c}{References for the BH
masses and the apparent magnitude of the host galaxies in $R-$band
used to estimate the BH masses in this work, of which (1) and (2)
are selected from Snellen et al. (1996) and \emph{NED} (NASA/IPAC
Extragalactic Database; http://nedwww.ipac.caltech.edu)
respectively.}
 \tablerefs{G09: Giroletti \& Polatidis 2009;
  V06: Vink et al. 2006; G05: Gugliucci et al. 2005; N09: Nagai et al. 2009;
  M99: Murgia et al. 1999; F02: Fanti et al. 2002; A1: this work assuming the upper limit of light speed;
  A93: Akujor et al. 1993; K97: Katz-Stone \& Rudnick 1997; A2: this work using the jet speed of 0.1$c$ in An et al. 2004;
  T01: Tyulbashev \& Chernikov 2001; T92: Taylor et al. 1992; K06: Kunert-Bajraszewska et al. 2006;
  L96: Lawrence et al. 1996; M03: Marziani et al. 2003; X99: Xu et al. 1999; P00: Peck et al. 2000;
  H01: Ho \& Peng 2001; X06: Xie et al. 2006; G04: Goncalves \& Serote 2004; H03: Hirst et al.
  2003; W09: Wu 2009 and references therein;  A00: Axon et al. 2000; W02: Woo \& Urry 2002;
   Z03: Zakamska et al. 2003.}

\end{flushleft}
\end{table*}

\end{document}